# Structured quantum search in NP-complete problems using the cumulative density of states

Keith Kastella and Richard Freeling
Keith.Kastella@Veridian.com
Freeling@ERIM-Int.com
*Veridan Ann Arbor Research and Development Center*
*P.O. Box 134008, Ann Arbor, MI 48114-4008*
(December 3, 2001)

In the multitarget Grover algorithm, we are given an unstructured $N$-element list of objects $S_i$ containing a $T$-element subset $\tau$ and function $f$, called an oracle, such that $f(S_i) = 1$ if $S_i \in \tau$, otherwise $f(S_i) = 0$. By using quantum parallelism, an element of $\tau$ can be retrieved in $O(\sqrt{N/T})$ steps, compared to $O(N/T)$ for any classical algorithm. In this paper, we show that in combinatorial optimization problems with $N = 4^M$, the density of states can be used in conjunction with the multitarget Grover algorithm to construct a sequence of oracles that structure the search so that the optimum state is obtained with certainty in $O(\log N)$ steps.

The last decade has seen great progress in the development of quantum algorithms that exploit quantum parallelism to solve problems much faster than classical algorithms. Perhaps the best known of these are the Shor[1] factorization and the Grover[2] search algorithms. Inspired by these results, a number of researchers have explored application of these methods to NP-complete problems such as decision and combinatorial search. Stimulated by the realization that for a completely unstructured problem, the Grover algorithm is optimal, most of this work has centered on the exploitation of problem structure combined with the Grover search[3,4]. Chen[5] et al assume that a sequence of auxiliary oracles can be constructed that mark states that agree with the target state in the first $2j$ bits. Then they use a recursive argument to show that $O(\log N)$ steps are required to find the target state.

This paper presents an algorithm that a) uses the density of states to structure quantum search efficiently in a combinatorial optimization problem and b) exhibits non-recursive construction of the unitary transformations required to implement it. In statistical physics, the density of states characterizes the number of continuum or discrete states as a function of the system energy. In combinatorial optimization, this generalizes naturally to the number of configurations as a function of a cost parameter to be optimized. For a system with $N$ states labeled $S_1, S_2, \ldots, S_N$, the cost of state $S_j$ is the real-valued function $C(S_j)$. Defining
$Q_c = \{S_j \mid C(S_j) \leq c\}$, the cumulative density of states is defined here as $\nu(c) = |Q_c|$, the number of elements of $Q_c$. While our goal is to find the unique state that minimizes $C(S)$, the density of states provides information about the distribution of costs but not about the cost of individual states. This is the type of information that is likely to be available to help structure quantum optimization problems. While not rigorously established, many combinatorial optimization problems are thought to be "self-averaging"[6]. This means that many properties of such problems converge to sample independent values in the limit of large $N$. For example, the traveling salesman problem is to find the shortest closed connected path connecting $M$ cities, given their pair-wise distance matrix. In this case there are $N = (M-1)!/2$ unique tours corresponding to states of the system. Nevertheless, the optimal tour distance, the system entropy and related quantities can be obtained analytically in the large-$N$ limit[7].

Our algorithm builds on the Grover search algorithm. To briefly review the Grover algorithm and define notation, we are given an $N$-element list of objects $S_i, i = 1, \ldots, N$. Within the list, there is a $T$-element subset $\tau$ and function $f$, called an oracle, such the $f(S_i) = 1$ if $S_i \in \tau$, otherwise $f(S_i) = 0$. The elements of $\tau$ are referred to as "good" or "target" states. The remaining elements are "bad". Let $P$ denote the operator with matrix elements $\langle S_i | P | S_j \rangle = 1$. The operator $U \equiv (2P/N - 1)\exp(i\pi f)$, referred to as the Grover iterate, is unitary. The unitarity of $U$ follows from the observation that in the $S_i$ basis, $P$ is an $N \times N$ matrix of ones with $P^2 = NP$ so $U^\dagger U = (2P/N - 1)^2 = 1$ (here $U^\dagger$ is the Hermitian conjugate). In the Grover algorithm, the system is prepared in the state $|\psi_0\rangle = \frac{1}{\sqrt{N}} \sum_i |S_i\rangle$ and $U$ is applied repeatedly. In most applications, $T \ll N$, and it turns out[11] that after $m \cong \frac{\pi}{4}\sqrt{\frac{N}{T}}$ iterations of $U$ the amplitude of the system is nearly $\frac{1}{\sqrt{T}}$ in the good states and very small in the bad states. In the usual application of the Grover algorithm, the system state is now measured. The outcome of the measurement has nearly unit probability to be in a good state, so that in this way, one of the good states is located, with some small probability of error.

While the original Grover algorithm contains a small probability of error, there have been several variants developed that can provide a good state with certainty[8,9,10]. One interesting such case occurs when $T = N/4$, in which case, *the solution is found with certainty in only a single iteration*[11] *(i.e., one application of $U$)*. The important point for our optimization application is that when $T = N/4$, one application of $U$ places all of the system amplitude uniformly in the target states while the bad state amplitudes vanish.

The other important building block in the algorithm presented below is the proof[12] that a combination of one- and two-bit quantum gates is universal in the sense that any unitary operator on $n$- qubits can be constructed using a finite number of two-qubit gates. Thus, for the algorithm to be efficiently implementable, it suffices to show that it is embodied as a unitary transformation. Then the procedure of Reck[13] *et al.* can be used to obtain an explicit decomposition of the transformation.

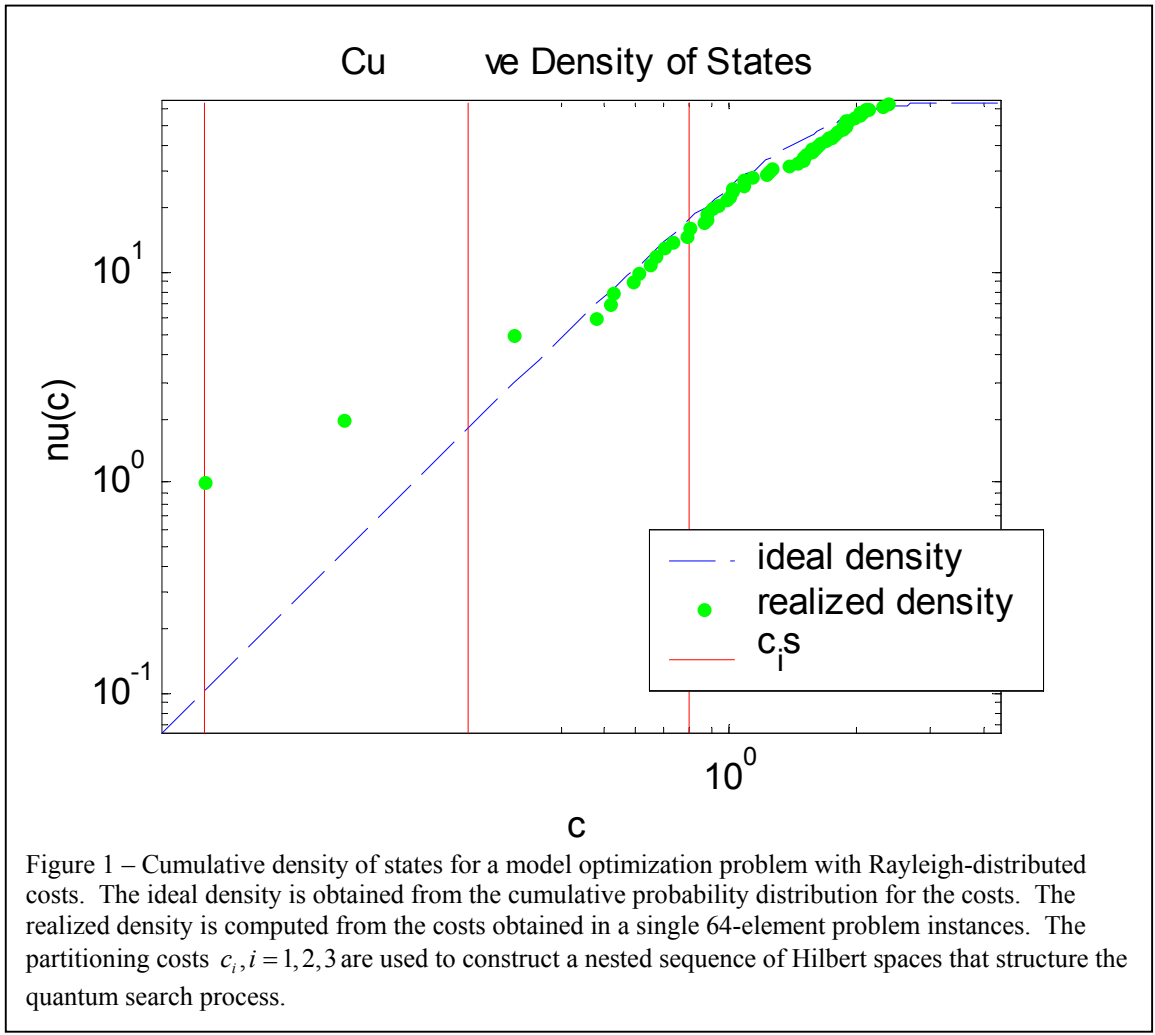

Figure 1 – Cumulative density of states for a model optimization problem with Rayleigh-distributed costs. The ideal density is obtained from the cumulative probability distribution for the costs. The realized density is computed from the costs obtained in a single 64-element problem instances. The partitioning costs $c_i, i = 1, 2, 3$ are used to construct a nested sequence of Hilbert spaces that structure the quantum search process.

Note that with perfect knowledge of the density of states, we also have the cost of the optimal state. This could in theory be used to construct an optimization algorithm by defining the oracle to be unity only on the optimal state. However, this would not be a very efficient algorithm, since it would require $O(\sqrt{N})$ iterations. As shown below, we can obtain $O(\log(N))$ complexity by extracting more information from the density of states.

To obtain this improved scaling, we use the cumulative density of states in two ways. First, it is used to construct a sequence of oracles that mark a nested sequence of good states containing the optimal state. Second, the density of states enables the construction of a sequence of projection operators that decouple the previous bad subspaces in subsequent processing. In this way the algorithm concentrates all of the amplitude on a progressively smaller portion of the Hilbert space. Once the amplitude has been forced to vanish on a bad subspace, it is held at zero during the remainder of the process.

The input assumptions on this optimization algorithm are as follows. Let a system have $N = 4^M$ states labeled $S_1, S_2, \ldots, S_N$. A real-valued cost function $C(S)$ is defined on the states. Our goal is to find the unique state that minimizes $C(S)$. In addition to the cost, we posit that the density of states $\nu(c)$ for this system is also provided, i.e., for any cost $c$ in the range of $C(S)$ we can compute number of states $\nu(c)$ such that $C(S) \leq c$.

The first step of the algorithm is to use $\nu(c)$ to construct a cost sequence $c_i, i = 1, \ldots, M$ such that $\nu(c_i) = 4^{M-i}$. Then define the sequence of oracles $f_0 = I$ and
$$f_i(S) = \Theta(c_i - C(S)), \quad i = 1, \ldots, M \quad (1)$$
where the $\Theta$ satisfies $\Theta(x) = 1, \ x \geq 0; \ 0$, otherwise. The functions $f_i$ are a set of real-valued linear operators on the Hilbert space $\mathcal{H}$ with basis $|S_i\rangle$. Written as a matrix, $f_i$ is $N \times N$ with $4^{M-i}$ ones on its diagonal and zeros elsewhere.

Recalling that $\langle S_i | P | S_j \rangle = 1$, define,
$$P_i = f_i P f_i, \quad (2)$$
$$D_i = \frac{2}{4^{M-i+1}} P_{i-1} - 1, \quad (3)$$
$$R_i = \exp(i\pi f_i), \quad (4)$$
$$V_i = f_{i-1} D_i R_i f_{i-1} + (1 - f_{i-1}), \quad (5)$$

The $V_i$'s are unitary, which can be seen as follows: $f_i^2 = f_i$ and $P_i^2 = f_i P f_i P f_i = 4^{M-i} P_i$, so $V_i^\dagger V_i = f_{i-1} D_i^2 f_{i-1} + (1 - f_{i-1}) = 1$. Initialize the system in the superposition state $|\psi_0\rangle = \frac{1}{2^M} \sum_j |S_j\rangle$ and define $|\psi_i\rangle = V_i|\psi_{i-1}\rangle$, $i = 1,\ldots,M$. Then measurement of the state $|\psi_M\rangle$ will produce the optimal state with unit probability.

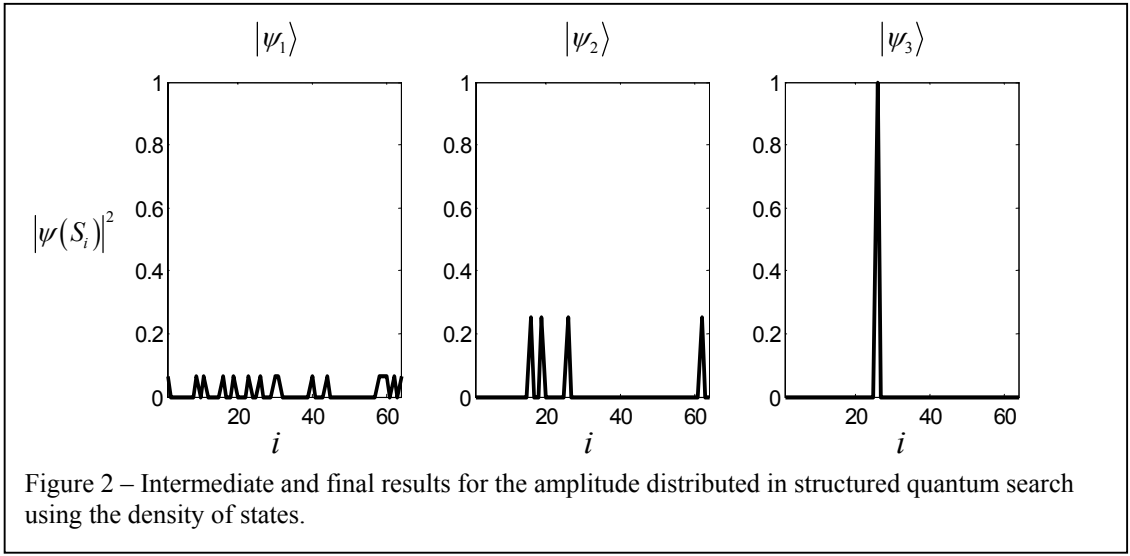

Figure 2 – Intermediate and final results for the amplitude distributed in structured quantum search using the density of states.

The optimality of $|\psi_M\rangle$ follows from the following observations. Denote the range of $f_{i-1}$ by $\mathcal{R}_{i-1} \subset \mathcal{H}$ has dimension $T_{i-1}$. Its compliment $\bar{\mathcal{R}}_{i-1} = \mathcal{H} \setminus \mathcal{R}_{i-1}$ is invariant under $V_i$. On $\mathcal{R}_{i-1}$, $D_i R_i$ is the usual Grover iterate, acting on the number $T_i = \nu(c_i) = 4^{M-i}$ of good states. The ratio of good states to non-invariant bad states is $T_i / T_{i-1} = 1/4$, so that the Grover iterate provides certain convergence after a single iteration. The net result is that $V_i$ moves the probability amplitude from $\mathcal{R}_{i-1}$ to $\mathcal{R}_i$ with certainty, while the amplitude in $\bar{\mathcal{R}}_{i-1}$ is held fixed at 0. At the conclusion of $M$ steps, all of the amplitude is concentrated in the single optimal state.

Figures 1 and 2 show this algorithm applied to a test problem with $N = 64$ states, indexed $i = 1,\ldots,64$. The costs are a random draw from a mean 1 Rayleigh distribution, $p(c) = c\exp(-c^2/2)$. Figure 1 shows a realization of the cumulative density of states $\nu(c)$. For comparison, we show the ideal cumulative density of states

$$\nu_{\text{ideal}}(c) \equiv N\int_0^c xdx \exp(-x^2/2)$$
$$= N\left(1 - \exp(-c^2/2)\right).$$ Figure 1 also shows the partitioning costs $c_i$ determined from $\nu(c)$ that are used to define the sequence of oracles (Eq. (1)). Figure 2 shows how the modulus of $|\psi_i\rangle$ evolves for the problem instance of Figure 1. Note that the number of computational basis states in $|\psi_i\rangle$ with non-vanishing amplitude is reduced by a factor of 4 with each step. The final amplitude is 1 in the optimal state ($i = 26$ in this instance), as desired.

In general applications the density of states is unlikely to be known with certainty. We have performed some anecdotal exploration of the algorithm behavior when there is some error in the density of states so that at each stage $\nu(c_i) \approx 4^{M-i}$ is only approximately correct. Interestingly, in this case the optimal solution *is still obtained with high probability, although certain convergence is lost*.

A likely more viable approach when $\nu(c_i) \approx 4^{M-i}$ is to use the approximate partitioning sequence of $c_i$'s to generate nested subsets, and then use quantum counting[14] to provide the exact values of the $\nu(c_i)$. To estimate $\nu(c_i)$ to $O(1)$ accuracy will require a counting register with roughly $O(\log N)$ bits. Once the $\nu(c_i)$ have been determined, we proceed as before with a slight modification to the iterate, which is required to guarantee convergence when the ratio of good states to bad states is no longer exactly ¼. As long as the ratio is close to ¼, only 1 or 2 iterations are required to provide assured convergence, and the complexity of the optimization process is still only $O(\log N)$. Each $\nu(c_i)$ is evaluated in a counting process that requires about $O\left((\log N)^2\right)$ work with a failure rate that can be made vanishingly small. The number of $c_i$ is held fixed at $O(\log N)$, so that the total complexity of the counting process is $O\left((\log N)^3\right)$.

In summary, we have shown that for combinatorial optimization problem with $N = 4^M$ configurations and a known cumulative density of states, this quantum search algorithm obtains the optimal configuration in $O(\log N)$ steps with unit probability. This new quantum search methodology appears to have potential for widespread applications in areas such as scheduling, bioinformatics, communications, and signal processing.

**ACKNOWLEDGEMENTS**

The authors benefited from stimulating discussion with K. Augustyn, I. Markov and C. Zalka. This work was supported by the Veridian Ann Arbor Research and Development Center and by the Air Force Research Laboratory and Air Force Office of Scientific Research under contract SPO900-96-D-0080.

[1] Shor P W, "Polynomial-time algorithms for prime factorization and discrete logarithms on a quantum computer," *SIAM J. Comp.,* 26, No. 5, pp 1484-1509, Oct. 1997.

[2] Grover L K, "Quantum mechanics helps in searching for a needle in a haystack," Phys. Rev. Lett. 79, 325-328 1997 (quant-ph/9706033); "From Schrodinger's equation to the quantum search algorithm," *Am. J. Phys.* **69** (7) July 2001, pp. 769-777.

[3] Cerf N J, Grover L K, and Williams CP, "Nested quantum search and NP-complete problems," quant-ph/9806078

[4] Hogg T, "Highly structured searches with quantum computes," *Phys. Rev. Lett.* **80**, 11 (1998) pp. 2473-2476.

[5] Chen G, and Diao Z, "Exponentially fast quantum search algorithm", quant-ph/0011109

[6] Mezard M, Parisi G, and Virasoro M, *Spin Glass Theory and Beyond*, (World Scientific, Singapore 1987)

[7] Mezard M, and Parisi G, "A replica analysis of the traveling salesman problem," *J. Physique* **47**, (1986) 1285-1296.

[8] Brassard G, Hoyer P, Mosca M and Tapp A, "Quantum amplitude amplification and estimation," quant-ph/0005055, (2000)

[9] Hoyer P, "On arbitrary phases in quantum amplitude amplification," *Phys. Rev.* **A62**, 052304 (2001), (quant-ph/0006031)

[10] Long G L, "Grover algorithm with zero theoretical failure rate," quant-ph/0106071 (2001)

[11] Boyer M, Brassard, G, Hoyer P and Tapp A, "Tight bounds on quantum searching," quant-ph/9605034, (1996)

[12] Barenco A, Bennett, C H, Cleve R, DiVencenzo D P, Margolus N, Shor P, Sleator T, Smolin J A and Weinfurter H, "Elementary gates for quantum computation," *Phys. Rev. A* **52**, 3457-3467, 1995 (quant-ph/9503016)

[13] M. Reck, A. Zeilinger, H. J. Bernstein, P. Bertani, "Experimental realization of any discrete unitary operator," *Phys. Rev. Lett.* **73**, 58 (1994)

[14] Brassard G, Hoyer P and Tapp A, "Quantum counting", quant-ph/9805082, (1998)